\documentclass[10pt,letterpaper,twocolumn]{article} 
\usepackage[margin=1in]{geometry}
\usepackage{ol2}
\usepackage[draft]{hyperref}
\usepackage{amssymb,latexsym,fontenc,yhmath,amsmath,graphics,subfigure,amsthm,mathtext,graphicx,lipsum,cuted,multicol}

\usepackage{lipsum}
\usepackage{fancyhdr}
\begin{document}

\twocolumn[ 

\title{Supervised classification-based stock prediction and portfolio optimization}



\author{Sercan Ar{\i}k$^{*,1}$, Bur\c{c} Ery{\i}lmaz$^{*,2}$, and Adam Goldberg$^{*,3}$}

\address{$^1$soarik@stanford.edu, $^2$eryilmaz@stanford.edu, and $^3$agoldberg@cs.stanford.edu }
\address{$^{1,2}$Department of Electrical Engineering, Stanford University, Stanford, CA 94305, $^3$Department of Computer Science, Stanford University, Stanford, CA 94305 }

\vspace{7 mm}

 ] 
\let\thefootnote\relax\footnotetext{$^*$Equal contribution}

\headheight 10pt 
  
\rhead{JUNE 2, 2014}
\section{Introduction}

As the number of publicly traded companies as well as the amount of their
financial data grows rapidly and improvements in hardware infrastructure and
information processing technologies enable high-speed processing of large
amounts of data, it is highly desired to have tracking, analysis, and
eventually stock selections automated. Machine learning has already attained an important place in trading and finance. One currently major area is high-frequency trading. There are many techniques in the literature and applications to predict short-term movements based on different stochastic models of temporal variations of stock prices \cite{HFT1},\cite{HFT2},\cite{HFT3} . Such approaches generally rely on treating individual stock data as a time series without analyzing correlations and patterns between different companies, mainly because of the limitations of processing very large data sets at very high-speeds. Another major application is valuation of the market based on economic parameters \cite{IND1},\cite{IND2},\cite{IND3},\cite{IND4},\cite{IND5},\cite{IND6}. For most of these applications the amount of data processed is limited and most do not drill down to the granularity of individual companies \cite{IND1},\cite{IND2},\cite{IND3},\cite{IND4},\cite{IND5},\cite{IND6}. Few works which studied portfolio optimization on individual company level focused on very small number of financial parameters \cite{SNV1}. 

A very large portion of the finance industry is comprised of mid- to long-term portfolio construction and is still mostly performed by hedge fund managers and financial analysts based on analysis and decision processes using company fundamentals. Considering the entire New York Stock Exchange (NYSE) stock market between the years 1993 and 2013 there are more than 27189 stocks (and that number grows almost everyday with new initial public offerings). Since this number is far larger than a human could manually handle, automation of financial analysis and investment decisions by modeling company fundamentals is a clear need.

In this project, we use machine learning techniques to address portfolio optimization. Our approaches are based on the supervision of prediction parameters using company fundamentals, time-series properties, and correlation information between different stocks. Rather than focusing on indicators of the overall economy, we focus on individual company data for the training phase. We reduce the problem to a classification problem since that makes it easier to evaluate our approach. We examine a variety of supervised learning techniques and find that using stock fundamentals is a useful approach for the classification problem, when combined with the high dimensional data handling capabilities of support vector machines. The portfolio our system suggests by predicting the behavior of stocks results in a 3$\%$ larger growth on average than the overall market within a 3-month time period.    

\section{Research Data}

\begin{figure}[h!]
\centerline{
\includegraphics[width=3.5in]{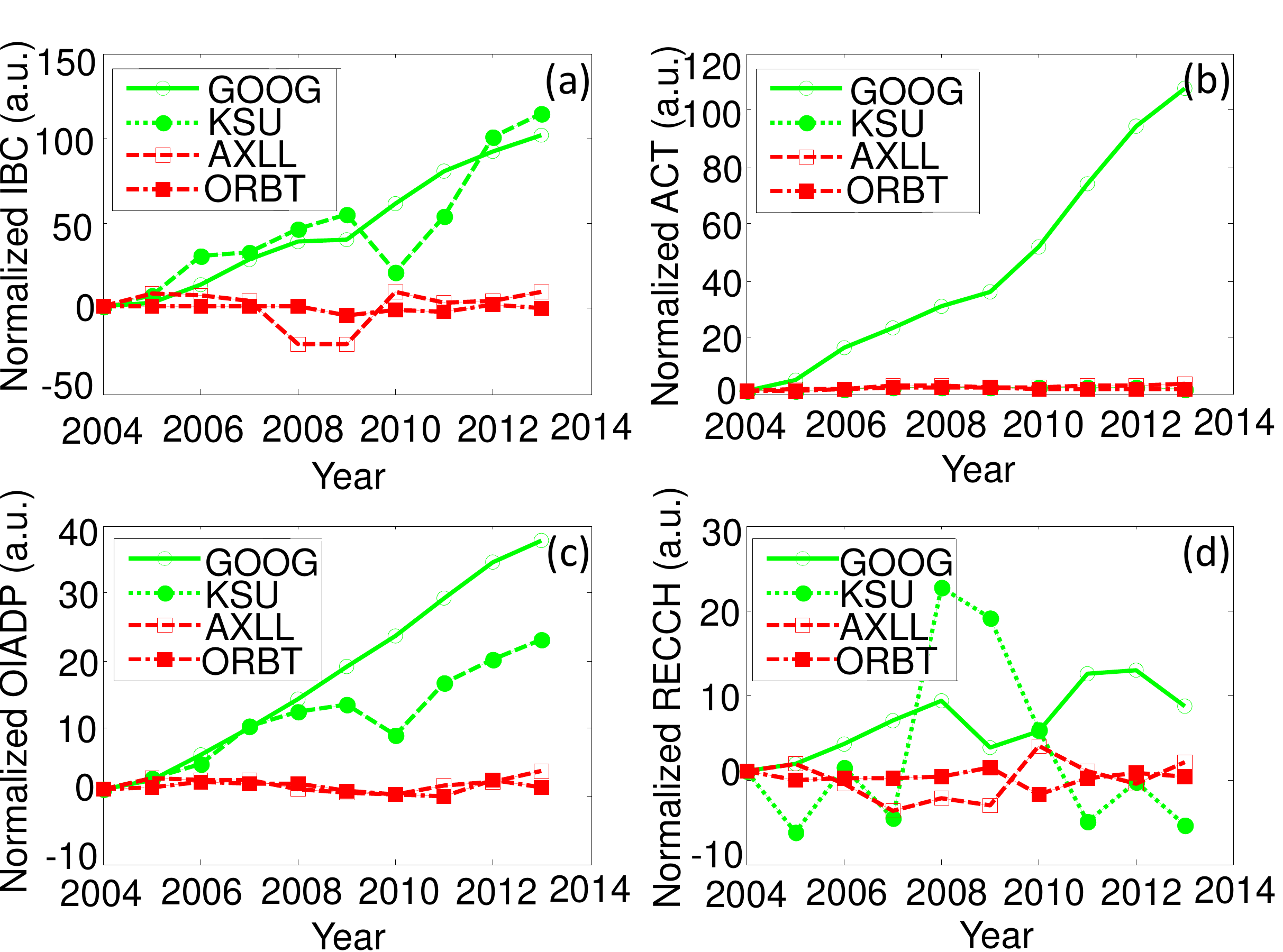}}
\caption{Example financial parameter values for two bullish stocks (GOOG and KSU), denoted with green circles, and two bearish stocks (AXLL and ORBT), denoted with red squares. Financial parameters represent (a) Income before extraordinary items, (b) Total current assets, (c) Operating Income After Depreciation (d) Accounts receivable/decrease.  }
\end{figure}

After a financial literature review and availability search, we pick 69 fundamental financial parameters to sufficiently represent the business fundamentals of each stock, as well as the general situation of the entire financial market and US economy. All of the financial parameters are taken with annual period, although their announcement date demonstrate almost uniform distribution throughout the year. As representative examples, four financial parameters (income before extraordinary items, total current assets, operating income after depreciation, and accounts receivable) among these are displayed in Fig. 1 for four stocks (GOOG, KSU, AXLL and ORBT) between the years 2004 and 2013. As formulated in Section 4, our classification approach is based on determining the annual performance of a stock compared to the average annual market performance (which is quantified with respect to the NYSE Composite Index). We define \emph{bullish} as `yielding positive gains with respect to market' and \emph{bearish} as `yielding negative gain with respect to market' and use them as our classification labels. In Fig. 1, there are examples of 2 bullish and 2 bearish stocks defined for the year 2013. Individual impact of various financial parameters might not be very discriminative when considered individually, however, as will be shown, financial parameter data becomes very useful when considered collectively and when combined with the flexibility and data handling capabilities of support vector machine classifier.

\begin{figure}[h!]
\centerline{
\includegraphics[width=3.4in]{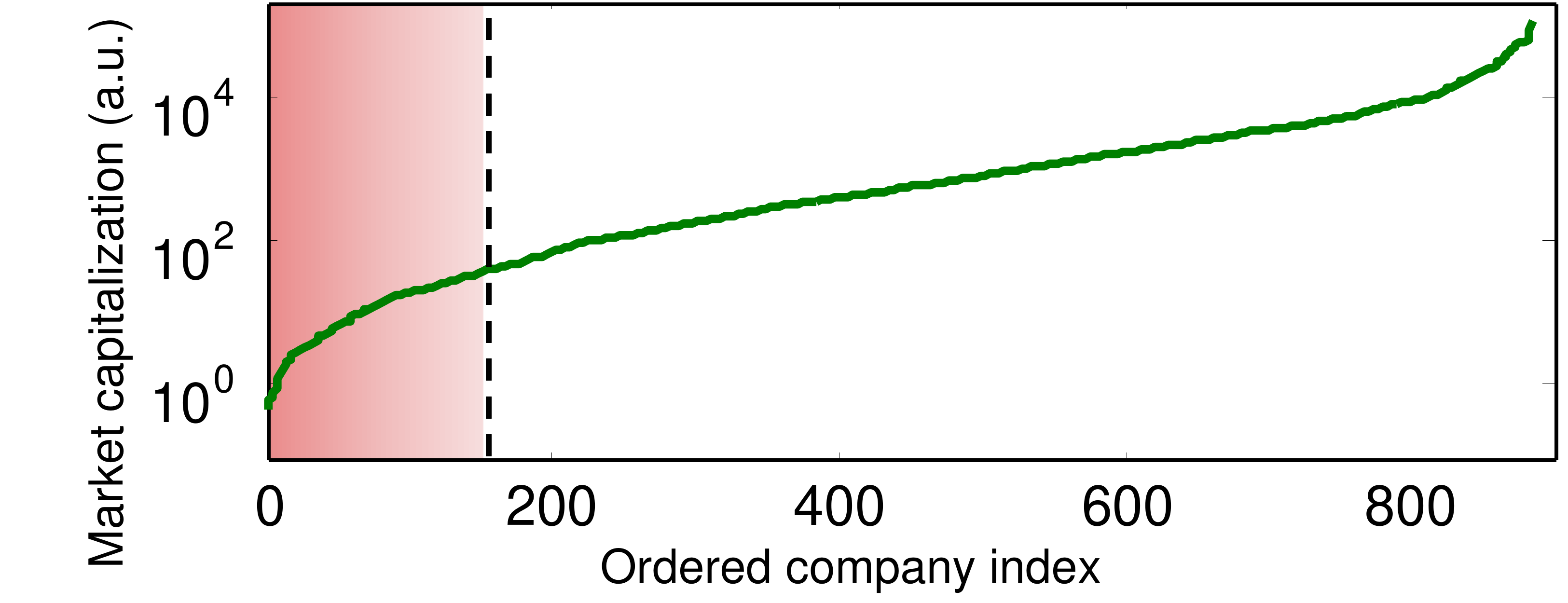}}
\caption{Market capitalization of the companies in the data set in ranked order. The companies shown in pink are discarded.}
\end{figure}

We initially consider all available NYSE stocks that have at least $50 \%$ of the analyzed financial parameters available. Among the entire set of stocks, we eliminate the ones that are not currently traded due to bankruptcy or acquisition. Since we need to be able to analyze temporal characteristics of stock behavior, we discard stocks with less than 10 years’ of financial data for the sake of completeness. Combined with the these limitations, we gather a data set comprising 1012 stocks. As can be observed from the market capitalization values in Fig. 2, there are more than four orders of magnitude difference between the financials of large and small companies. The behavior of the stock prices of extremely small companies typically do not have a pattern that can be modeled analytically. The trend of their stock prices are often dominated by speculations barely dependent on financial fundamentals. Hence, for more robust implementation of the supervised learning techniques, we further eliminate 152 companies with the smallest market capitalization.

\section{Data Preprocessing}

Given the large size ($ 69 \cdot 10 \cdot 860 = 593400$ data points in total) of the entire set of financial parameters, preprocessing is critical for the efficiency and accuracy of the learning and prediction techniques. In this section, we describe the data preprocessing approaches we use before application of supervised classification.

Firstly, the financial data set of 860 companies include missing and errorenous values for some financial paremeters. We address this problem in two steps. Firstly, we eliminate financial parameters with a fraction of missing data larger than a certain threshold (note that this is different from eliminating some stocks explained in Research Data section, where we eliminate stocks that are missing data more than a threshold). Conservatively setting this threshold to $5 \%$ for the years between 2004-2013, we eliminate 17 out of 69 features from our financial data set, and are left with 52 financial parameters. Secondly, we perform imputation by assigning the mean (averaged over all stocks within the given year) of that given financial parameter to the missing data entry, which we don't believe skews our feature distribution as few values are missing after our aforementioned company and feature selections. 

As the relative change of a financial parameter rather than the monetary value itself is more important for the performance, we normalize each parameter with respect to its value at the beginning of the financial data time frame (similar to Fig. 1). Then, we normalize each distribution to zero mean and unit variance for a given feature (across all stocks in the given year), since the supervised classification implementations we use are optimized for this normalization. 

\begin{figure}[h]
\centerline{
\includegraphics[width=3.4in]{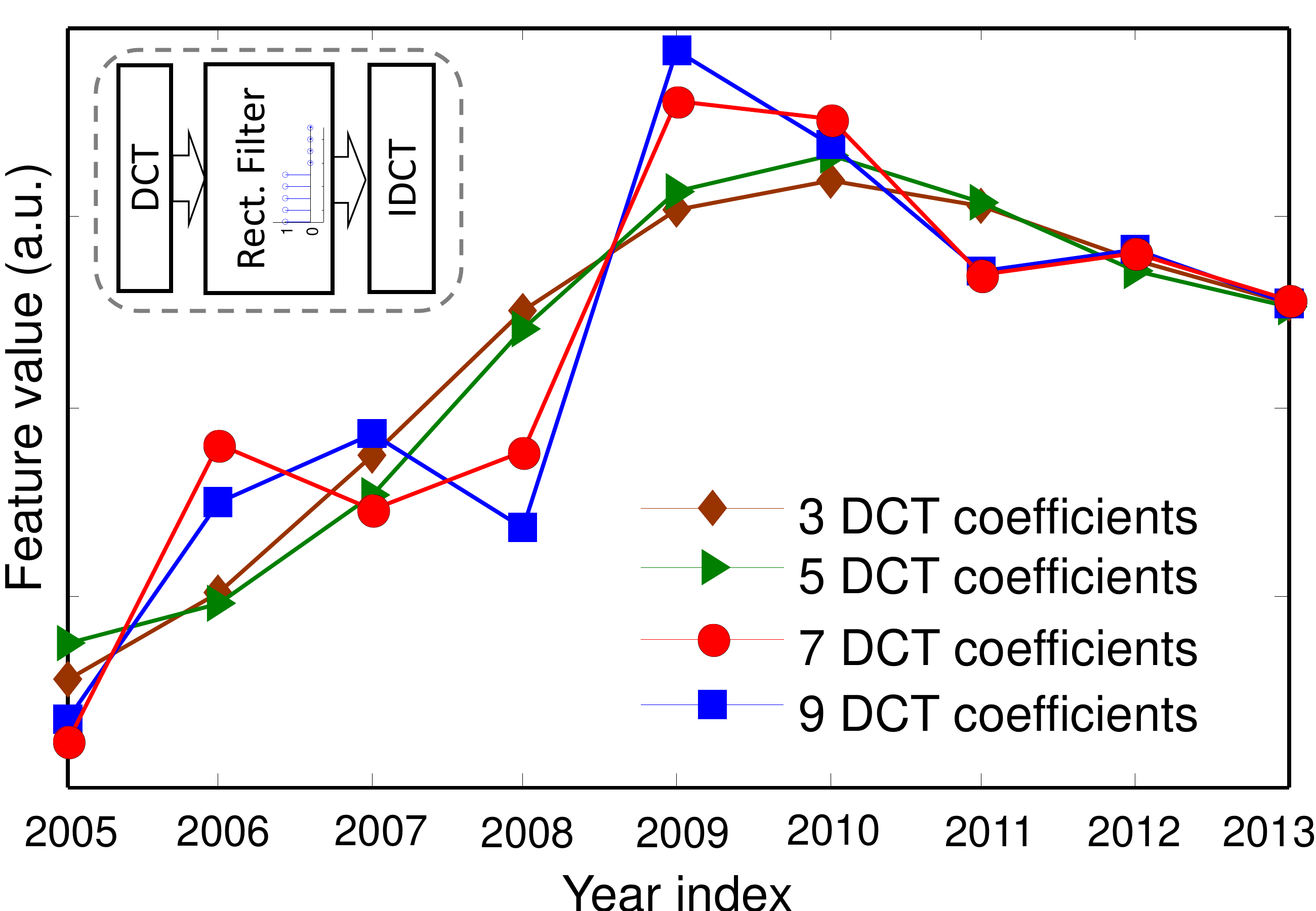}}
\caption{The feature value over time for different values of DCT coefficients. Inset: block diagram of the filtering operation.}
\end{figure}

For robustness of the classification, we consider two smoothing strategies. Most of the financial parameters do not have a smooth variation from one year to another, but rather have an abrupt change over the years (see Fig. 3). To filter out this high-frequency volatility, we apply filtering in the discrete cosine domain. We consider filtering using an ideal rectangular discrete filter which is equivalent to setting a particular number of high-order coefficients to zero in discrete cosine domain. The discrete cosine transform (DCT) of a discrete time-domain signal $f[n]$ ($n = 1,...,N$) is:
\begin{equation}
F[k] = \sqrt{\frac{2}{N}} (2-\delta[k]) \sum_{n=1}^{N}f[n] \cos\left ( \frac{\pi(2n-1)(k-1)}{2N} \right ),
\end{equation}
where ($k = 1,...,N$) and $\delta[k]$ is the Dirac delta function. The filtering operation with rectangular filter with width $h$ yields: 
\begin{equation}
\widetilde{F[k]} = \left\{\begin{matrix}
F[k] \; \; \; \; \; $for$ \; \; \;k = 1,...,h ,\\ 
0 \; \; \; $for$ \; \; \;k = h+1,...,N .
\end{matrix}\right.
\end{equation}
Finally, the filtered signal is transformed back into the time domain applying inverse DCT:
\begin{equation}
\widetilde{f[n]} = \sum_{k=1}^{N}\sqrt{\frac{2}{N}} (2-\delta[k]) \widetilde{F[k]} \cos\left ( \frac{\pi(2n-1)(k-1)}{2N} \right ).
\end{equation}
Fig. 3 exemplifies the impact of the rectangular discrete filter width $h$ (or the number of non-zero DCT coefficients) on smoothing. Based on our empirical observations, we choose a rectangular filter width of $7$ for all of the variables in our financial data set. Lastly, we consider smoothing based on principal component analysis. For robustness of the supervised classification, we eliminate the smallest $15 \%$ principal values, which are computed using singular value decomposition.   

\section{Stock Prediction with Supervised Learning}

The rationale of our stock prediction technique is based on determining classification parameters using the relationships between the past financial parameters and their impacts on the past performance. In this section, we first formulate the general approach and then specifically describe the techniques used in this paper.  

Let $k$ be the number of features. For a stock with index $i$ ($1 \leq i \leq n $), let the feature vector $\mathbf{x}^{(i)} [t] \in \mathbb{R}^k$ denote the corresponding financial parameters of year $t$ after pre-processing. Let the label $y^{(i)} [t+1] \in \left \{ -1,1 \right \}$ denote the relative market performance (i.e. either bearish or bullish) of the stock $i$ in a given period of time after the announcement date of financial parameters of year $t$ ($t \in \left \{ 2013, 2012, 2011 ... \right \}$). The goal is to predict $y^{(i)} [t]$ for each $i$ using the financial parameters of past $L$ years, i.e. based on the feature vector: 
\begin{equation}
\mathbf{X_{[t-1;t-L]}}^{(i)} = \begin{bmatrix}
\mathbf{x}^{(i)}[t-1] \\ 
\mathbf{x}^{(i)}[t-2] \\
\vdots \\
\mathbf{x}^{(i)}[t-L]
\end{bmatrix}.
\end{equation} 

For supervision of the learning parameters, known performances of all of the stocks in the data set of past $M$ years are used. The corresponding training data matrix is formed as:
\begin{equation}
\mathbf{X_T} = \begin{bmatrix}
\mathbf{X_T}^{(1)}\\ 
\mathbf{X_T}^{(2)}\\
\vdots \\
\mathbf{X_T}^{(n)}
\end{bmatrix},
\end{equation} 
where 
\begin{equation}
\mathbf{X_T}^{(i)} = \begin{bmatrix}
\mathbf{X_{[t-2;t-L-1]}}^{(i)}\\ 
\mathbf{X_{[t-1;t-L-2]}}^{(i)}\\
\vdots \\
\mathbf{X_{[t-M-1;t-L-M]}}^{(i)}
\end{bmatrix}.
\end{equation} 
And the vector of corresponding known labels (determined by the relative market performance in that given time frame) of past $M$ years are:
\begin{equation}
\mathbf{Y_T} = \begin{bmatrix}
y^{(1)} [t-1] \\ 
y^{(1)} [t-2] \\
\vdots \\
y^{(1)} [t-M] \\
\vdots \\
y^{(n)} [t-1] \\ 
y^{(n)} [t-2] \\
\vdots \\
y^{(n)} [t-M] \\
\end{bmatrix}.
\end{equation} 

After applying all of the pre-processing operations, we end up with $k = 52$ features as explained in the previous section. Our observations suggest that both financial parameters and stock performance in the past beyond a particular time frame have negligible effect on the current performance. Hence, we make choices of $M = 5$ years and $L = 5$ years. To further narrow down the problem, we limit the performance metric to the time frame of 3 months after the announcement date of financial parameters, e.g. $y^{(i)} [2013] = 1$ if the stock $i$ outperforms the market at the end of the three months after the accouncement of financials of the year 2012. 

\begin{figure}[h]
\centerline{
\includegraphics[width=2.5in]{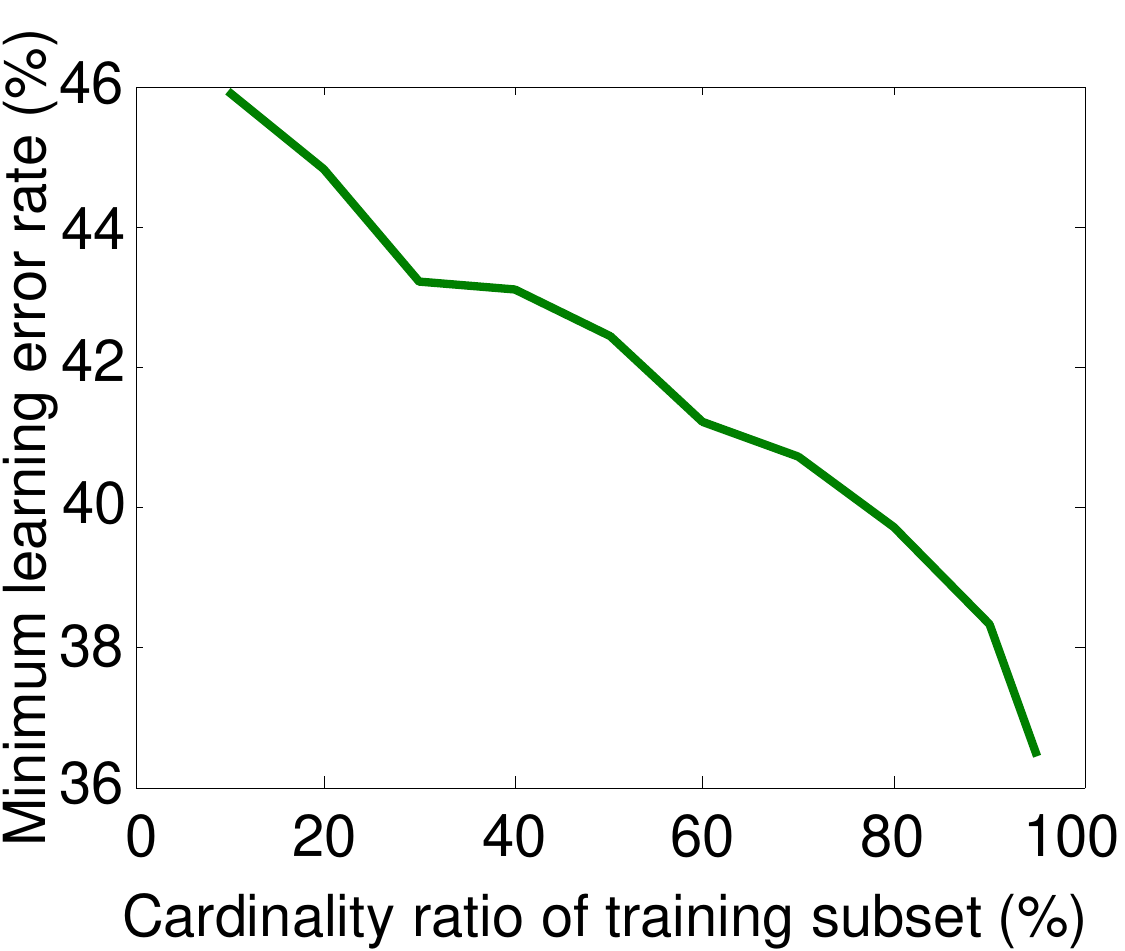}}
\caption{Minimum learning error rate vs. cardinality ratio of training subset (obtained using SVM classification as explained in the test).}
\end{figure}

We partition the entire training data set of size $M \cdot n$ into two subsets and partition $\mathbf{X_T}$ and $\mathbf{Y_T}$ accordingly. We use the first subset to determine the classification parameters and the second subset to test the training accuracy of the classification obtained by the first data set. This also allows us to consider different subsets of the training set to optimize the learning performance by reducing the impact of speculative companies or abrupt changes on supervision. Fig. 4 shows the the impact of the cardinality ratio of the first subset on the training error rate. To have low enough prediction error rate while having a reasonable number of elements in the testing data set, we fix this cardinality ratio to $90 \%$ for the rest of the results. Since consideration of all combinations with the given cardinality ratio have prohibitively high computational complexity, we only consider a particular number of random realizations for training. 

We initially consider supervised classification using Decision Trees, $k$-Nearest Neighbors, Naive Bayes and support vector machine (SVM) techniques. Our results suggest that SVM outperforms the other three classification techniques with more than $3 \%$ prediction accuracy, provides more robustness to the changes in the training data set, and is more computationally efficient for large data sets. Therefore, in this paper we focus on stock prediction using SVM based classification. Other studies \cite{IND1},\cite{IND3},\cite{IND4},\cite{IND6}  also show SVM is well-suited for financial applications with its capability to model high dimensional feature spaces since financial parameters can take very different functional forms (see Fig. 1 for example). 

We perform optimization of several SVM parameters for the highest prediction accuracy by exponential grid search. We choose the Gaussian kernel based on the motivation explained above, that is also motivated in the literature to work well for problems involving time series analyses \cite{RBF}. We choose a box constraint parameter of 0.8. We allow up to $10 \%$ of the support vectors to violate Karush-Kuhn-Tucker (KKT) conditions considering the nature of the problem such that lower prediction accuracy values are acceptable. Because of the convergence issues of commonly used sequential minimal optimization based updates, least-squares based parameter updates are applied.

\section{Results}

\begin{figure}[h]
\centerline{
\includegraphics[width=3.6in]{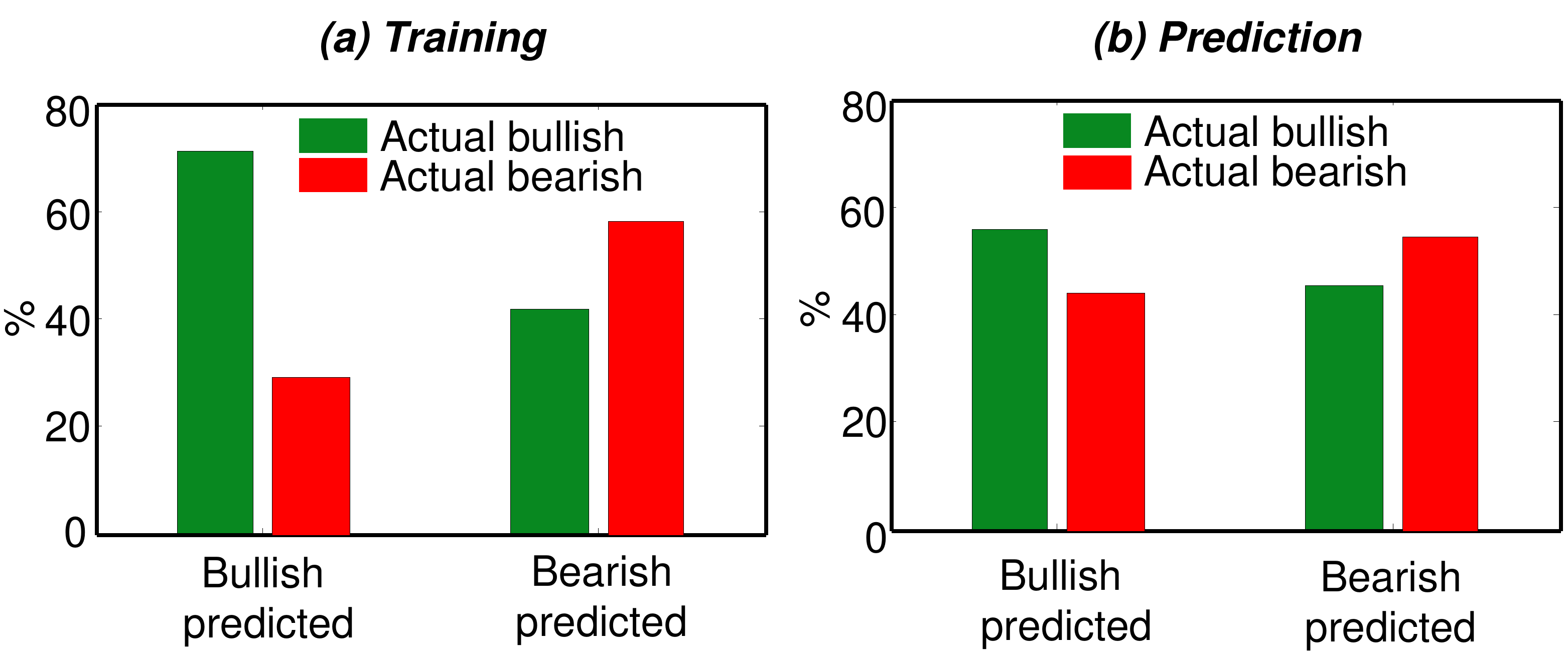}}
\caption{Average confusion matrix values (bars in each plot show the ratios of true positive, false positive, false negative, true negative respectively) in (a) Training and (b) Prediction.}
\end{figure}

Fig. 5 depicts the accuracy of the classification in both training (for the partitioned data set as explained in previous section) and prediction (for the entire data set corresponding to year 2013). Error values are shown separately for bullish and bearish labeled stocks, averaged over 100 random realizations of training and prediction. In training, the average true classification ratios are $71.2 \%$ for bullish stocks and $60.2 \%$ for bearish stocks; and in prediction the average true classification ratios are $58.8 \%$ for bullish stocks and $57.9 \%$ for bearish stocks. Moreover, for all random realizations the prediction accuracy is found to be greater than $56.2 \%$, suggesting the robustness of the technique even more noisy stocks are included in the training data set.

\begin{figure}[h]
\centerline{
\includegraphics[width=3.6in]{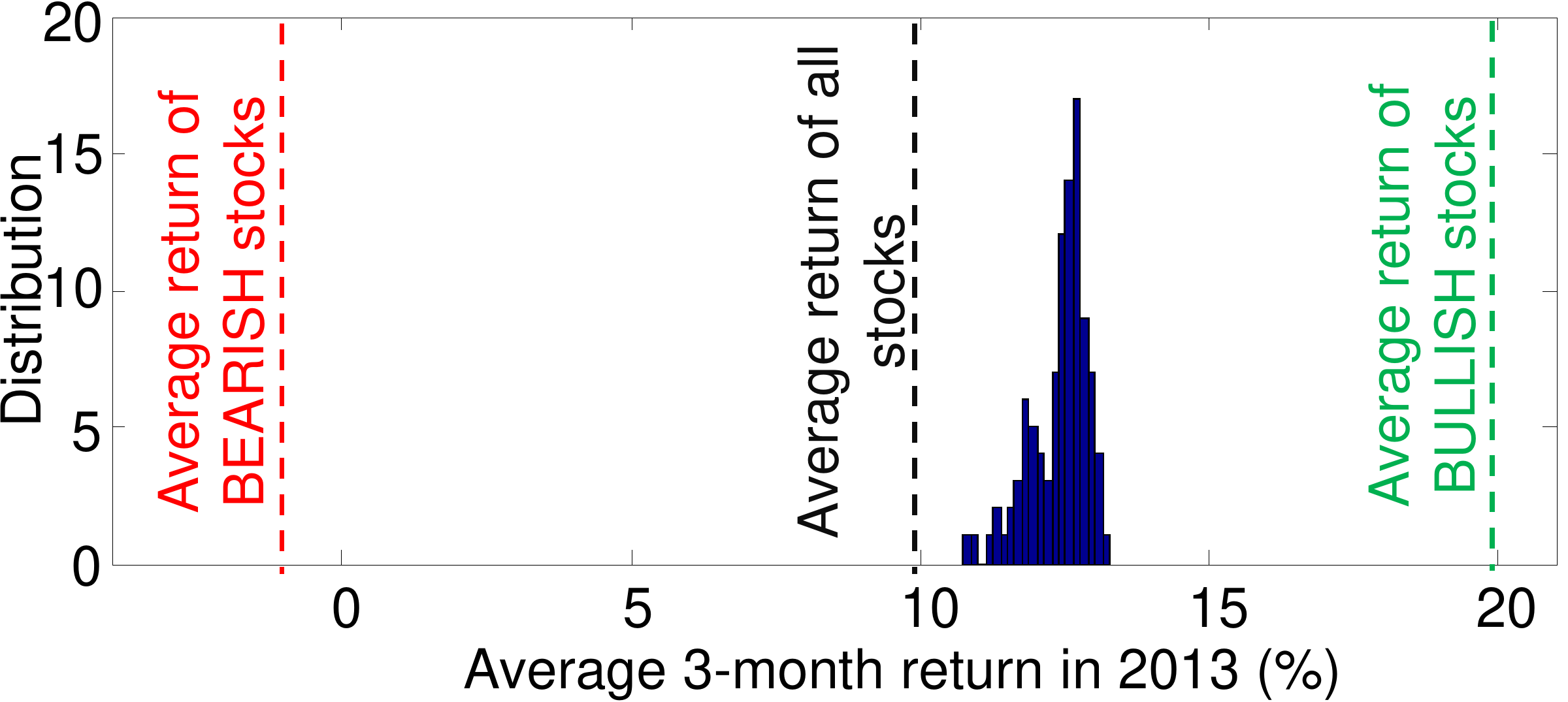}}
\caption{Distribution of average 3-month return of the portfolio obtained by uniform weighing of bullish labeled stocks (shown with bars), average return of all stocks (black dashed lines), average return of all bearish stocks corresponding to $0 \%$ prediction accuracy (red dashed line), and average return of all bullish stocks corresponding to $100 \%$ prediction accuracy (green dashed line).}
\end{figure}

As a more meaningful metric, we evaluate the performance of the portfolio comprised of the bullish classified companies by the proposed classification technique. We simply assume an equal share allocation among the bullish classified companies. (Share allocation optimization can yield higher returns and is further discussed in Future Work.) Fig. 6 shows the distribution of the returns of the portfolio obtained over 100 random relatizations, as well as the average return of all stocks and average returns of actual bullish and bearish stocks. As can be observed, market outperformance is obtained for all of the random realizations. The average 3-month return of the the optimized portfolio is approximately $3 \%$ higher than the market average.

\section{Conclusions and Future Work}

In this paper, we realize an supervised classification based predictor for mid- to long-term portfolio construction. Since our analyses suggest that SVMs give better performance over other supervised learning techniques (such as Naive Bayes and k-Nearest Neighbors), to minimize prediction error rate, we focus on SVMs to further optimize our classifier. Having determined a set of financial parameters after an extensive literature review, we develop data mining techniques to effectively collect them for the time period of interest. We apply state-of-the-art statistical data analysis tools (discrete cosine transform and principal component analysis) to smooth the information content of the stock fundamentals data and reduce dimensionality. We find that using several financial parameters for all the companies in the dataset collectively while utilizing the high dimensional feature space handling capabilities of SVMs is very effective to discriminate between bullish and bearish stock growth for portfolio optimization. To further increase the success rate of prediction of our classifier, we choose optimal values for several SVM parameters. The portfolio of companies predicted to be bullish by the classifier yields $3 \%$ more return on average over a 3-month period than the NYSE composite index. This is a significant step towards realizing an automated stock picking algorithm using the principles of artificial intelligence and machine learning.

We believe there are important fundamental open research problems and room for improvement in this area. From a machine learning perspective, it will be useful to study the performance of our classifier for a broader range of different kernels and optimization techniques such as Adaptive Simulated Annealing and the Fourier kernel \cite{RBF} . This will allow us to see any overfitting or underfitting issues that might be occurring in our classifier. In addition, we can try a modified version of SVM as suggested in \cite{SUG1} where the data closer to the present time is assigned more importance. From a finance perspective, the empirical distributions of financial parameters and their impact on stock performance are actively studied and hinder the application of machine learning techniques based on \emph{a priori} statistics. The portfolio we build in this work assigns equal weight for each stock predicted to be bullish. An important future goal is to develop a model to set a price target for the chosen stocks in the portfolio, that will allow optimization of the amount of each stock in the portfolio. In order to make this work more practical and application specific, we can also construct a portfolio based on a risk parameter according to user needs, and determining the weights of each stocks in the portfolio accordingly. Correspondingly, we are currently examining various regression techniques that can be generalized for our problem, as well as hybrid classification and regression approaches. We are also investigating efficient modeling of the temporal correlations of the financial parameters in order to effectively model the impact of the past years' data to improve learning. Potential approaches we are considering are spectral classification based on fast Fourier transforms or optimized filtering techniques.

\end{document}